\documentclass[aps,
twocolumn,
showpacs,
prl,
amsmath,amssymb,floatfix,superscriptaddress]{revtex4-2}
\usepackage{blindtext}
\usepackage{graphicx}
\usepackage{bm}
\usepackage{hyperref}
\usepackage{amsmath}
\usepackage{mathtools}

\usepackage{pdfpages} 
\usepackage{pgffor} 

\makeatletter
\AtBeginDocument{\let\LS@rot\@undefined}
\makeatother

\pdfximage{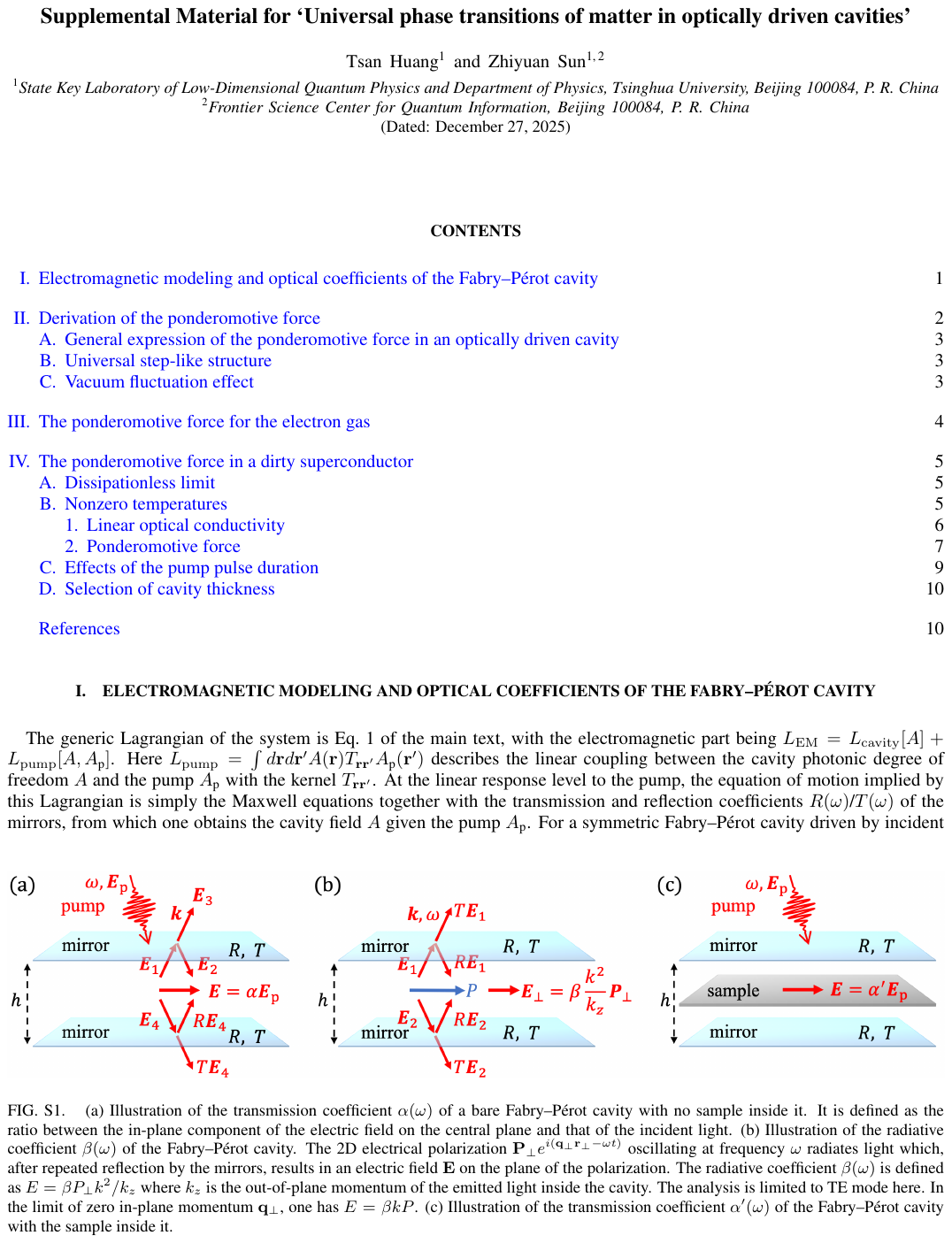}
\def\numbersupplementpages{\the\pdflastximagepages}

\newif\ifarXiv
\arXivtrue 

\usepackage{cleveref}

\usepackage{hyperref}
\hypersetup{%
	pdftitle={cavity}, %
	pdfauthor={Zhiyuan Sun},%
	pdfpagemode={UseNone},%
	pdfstartview={FitH},%
	breaklinks=true,%
	citecolor=blue,%
	colorlinks=true,%
	linkcolor=blue,%
	urlcolor=blue
}

\newcommand{\unit}[1]{\,\mathrm{#1}} 
\newcommand{\equa}[1]{Eq.~\eqref{#1}} 
\newcommand{\fig}[1]{Fig.~\ref{#1}}

\newcommand{\rom}[1]{\uppercase\expandafter{\romannumeral #1\relax}}

\usepackage[T1]{fontenc}
\usepackage{times}{\Large}

\graphicspath{{figures/}}

\begin{document}
\title{Universal Phase Transitions of Matter in  Optically Driven Cavities}


\author{Tsan Huang}
\affiliation{State Key Laboratory of Low-Dimensional Quantum Physics and Department of Physics, Tsinghua University, Beijing 100084, P. R. China}

\author{Zhiyuan Sun} \email{zysun@tsinghua.edu.cn}
\affiliation{State Key Laboratory of Low-Dimensional Quantum Physics and Department of Physics, Tsinghua University, Beijing 100084, P. R. China}
\affiliation{Frontier Science Center for Quantum Information, Beijing 100084, P. R. China}

\begin{abstract}		
Optical cavities have been widely applied to manipulate the properties of solid state materials inside it. 
We propose that in  systems embedded within optical cavities driven by incident pump light, the pump induces generic phase transitions into new non-equilibrium  steady states.
This effect arises from the ponderomotive potential, the effective static potential exerted by the pump on the low energy degrees of freedom, which exhibits a universal step-like structure that pushes the matter degrees of freedom in the direction that red-shifts the  cavity photon modes. 
For a two dimensional electron liquid in a driven cavity, this  step-like potential pushes the electron density to jump to a smaller value so that a hybrid cavity photon mode is red shifted to  slightly below the pump frequency.
Similarly, for a dirty superconductor in such a driven cavity, this potential acts on the superconducting order parameter and causes a first order phase transition to a new steady state with a smaller gap.
By realistic electromagnetic modeling of the cavity that includes all cavity modes,
we construct the non-equilibrium  phase diagrams for experimentally relevant devices.


\end{abstract}

\maketitle


An emerging platform for tuning the properties of solid state materials is to couple them to optical cavities~\cite{Deng.2010, frisk2019ultrastrong,forn2019ultrastrong, xiong_perovskite_2021, hubener2021engineering, schlawin2022cavity, bloch_strongly_2022}, which enables precise control of light-matter interactions in confined geometries. 
Apart from the  hybridization between collective modes and cavity photons into polaritons~\cite{Deng.2010,xiong_perovskite_2021,
xiong_nonlinear_2022, Chen_Zhang_2024}, 
researchers are actively exploring the possibility that the cavity photon fluctuations may even manipulate phases of matter  such as  superconductivity~\cite{sentef2018cavity, schlawin2019cavity,chakraborty2021long,lu2024cavity,Kozin.2025,keren2025cavity}, ferroelectricity~\cite{ashida2020quantum,latini2021ferroelectric,curtis2023local}, 
excitonic order~\cite{mazza2019superradiant,andolina2019cavity},
magnetism~\cite{weber2023cavity,kass2024manybodyphotonblockadequantum}, 
metal-insulator transitions~\cite{jarc_cavity-mediated_2023}, superradiance~\cite{Kono.2025_superradiance}, and  topology~\cite{felice2022breakdown,Dmytruk2022, Rubio.2023,Faist.2024, enkner2024enhancedfractionalquantumhall, yang2025quantumhalleffectchiral}. 
Unfortunately, because of the smallness of the quantum and thermal fluctuations of cavity photons,  it is experimentally challenging to realize cavity engineering of equilibrium many-body phases~\cite{felice2022breakdown, jarc_cavity-mediated_2023, kim2024cavity, Faist.2024, enkner2024enhancedfractionalquantumhall, thomas2025exploring}.
On the other hand, one may greatly enhance the cavity photon fields by pumping it with incident light within optically driven cavities, so that the material inside it could be much more effectively controlled  in the \emph{non-equilibrium} regime~\cite{dalla2012keldysh, ritsch2013cold, curtis2019cavity, gao2020photoinduced, Rodriguez.2020, chiocchetta2021cavity, sentef2020quantum, juraschek2021cavity, Sun2024exciton,cheng2024reversible}.



\begin{figure}[t]
	\includegraphics[width=\linewidth]{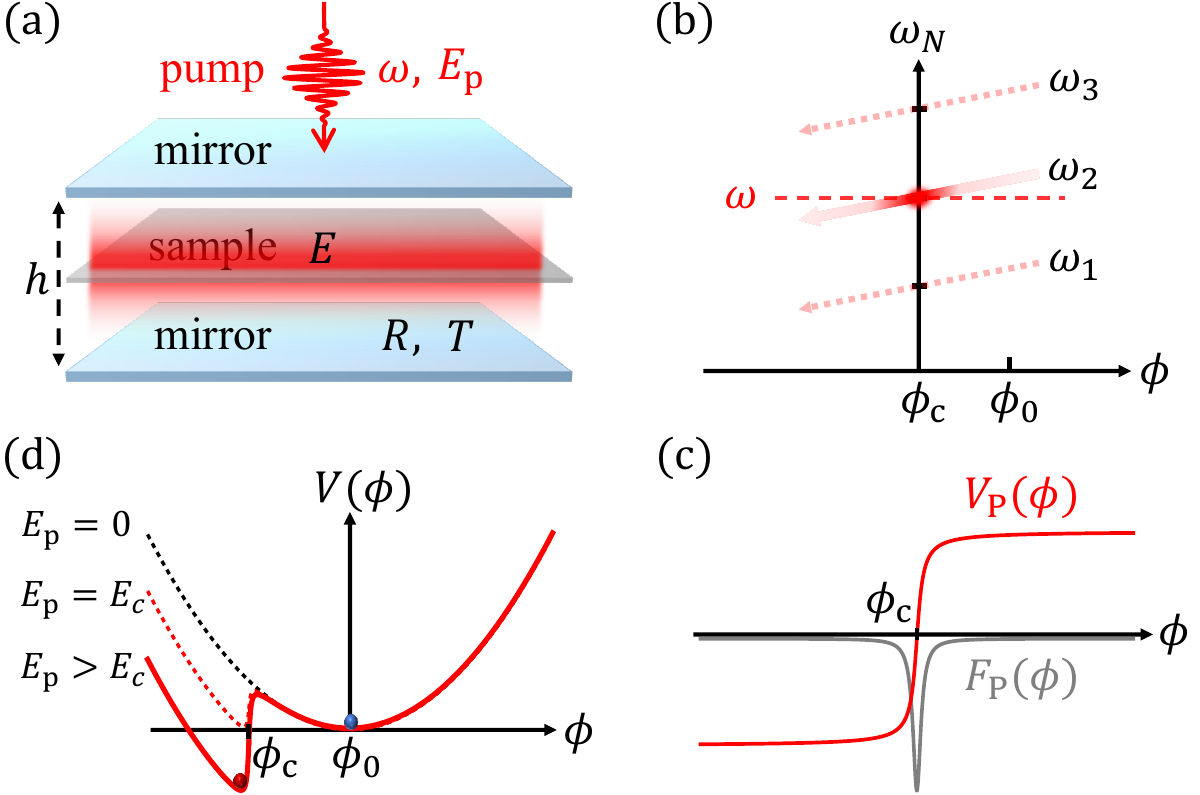} 
	
	\caption{(a) Schematic of a Fabry–Pérot cavity of thickness $h$ with the sample (an ultra-thin film or a strictly two-dimensional material) inside it. The mirror has the reflection and transmission coefficients $R(\omega)$ and $T(\omega)$. The red color scale represents the in-plane electric field of an anti-node cavity photon mode.  (b) The dressed cavity photon frequencies $\omega_{\text{N}}$ at zero in-plane momentum as functions of $\phi$, the slow degree of freedom on the sample. The red color scale represents the  ponderomotive force on $\phi$, which peaks at $\phi_{\text{c}}$ when the cavity photon is in resonance with the pump $\omega$.
    (c) The ponderomotive potential $V_{\text{P}}$ and ponderomotive force $F_{\text{P}}$ felt by $\phi$.
    (d) The typical total potential felt by $\phi$ for $E_{\text{p}}=0$ (black dashed line), $E_{\text{p}}= E_{\text{c}}$ (red dashed line), and $E_{\text{p}}>E_{\text{c}}$ (solid red line). }
	\label{fig:FPcavity} 
\end{figure}

In this letter, we investigate the  non-equilibrium steady states  of  many-body systems (the `sample' in \fig{fig:FPcavity}(a)) placed within optically driven Fabry–Pérot (F-P) cavities. 
By an electromagnetic approach   that analytically accounts for the coupling of matter to all cavity photon modes without artificial parameters, we uncover a universal step-like ponderomotive potential~\cite{sun2024floquet} on the sample
generated by the pump, leading to generic phase transitions for almost all systems.  Non-equilibrium phase diagrams are predicted for the two dimensional electron gas (2DEG) and dirty superconductors to be verified by experiments.

We consider a symmetric F-P cavity driven by the uniform pump light $\mathbf{E}_{\text{p}} e^{-i \omega  t}+c.c.=-\partial_{t} \mathbf{A}_{\text{p}}$ represented by the electromagnetic vector potential $\mathbf{A}_{\text{p}}=\mathbf{A}_{\text{p0}} e^{-i \omega  t}+c.c.$  with the frequency $\omega$, as shown in \fig{fig:FPcavity}(a). 
The sample inside it is modeled as a two dimensional (2D) plane representing a 2D thin film or an atomic layer.
The generic Lagrangian for such a system could be written as
\begin{eqnarray}\label{eq:initialL}
	L=L_{\text{M}}[X,\phi,\mathbf{A}]+L_{\mathrm{EM}}[\mathbf{A},\mathbf{A}_{\text{p}}]
\,.
\end{eqnarray}
Here $L_{\text{M}}$ is the Lagrangian for the matter degrees of freedom in the sample: the fast field $X$ and the slow field $\mathbf{\phi}$~\cite{sun2024floquet} which are coupled to the  vector potential $\mathbf{A}(\mathbf{r},t)$, the photonic degree of freedom inside the cavity.
The slow field $\phi$ represents the degrees of freedom whose dynamics is much slower than the driving frequency~\cite{sun2024floquet}, e.g.,
a soft lattice distortion in an insulator,
or the slow component of the order parameter in a superconductor. 
To grasp a physical picture, we encourage the readers to think of $\phi$ as  the electronic density $n$ of a metal in the following.
The fast field $X$ are the degrees of freedom that oscillate at the driving frequency or its higher harmonics, such as the high energy electronic transitions. 


The term
$L_{\mathrm{EM}}=L_{\mathrm{cavity}}+L_{\mathrm{pump}}$ contains the quadratic Lagrangian $L_{\mathrm{cavity}}$ for the cavity photon modes and $L_{\mathrm{pump}}$ for their  linear coupling to the external pump field $\mathbf{A}_{\text{p}}$. 
They lead to the Maxwell equations and the mirror's transmission/reflection coefficient $R(\omega)$/$T(\omega)$ that encode the information of all the cavity photons and their coupling to the pump.
It is convenient to define
the dimensionless coefficients:
\begin{align}\label{eq:alpha}
	\alpha' &=\frac{\alpha}{1-i\beta \sigma /c}
	,\quad      \alpha=\frac{Te^{i\frac{kh}{2}}}{1-Re^{ikh}}
\end{align} 
where $h$ is the cavity thickness,
$\alpha'(\omega)$ is the effective linear transmission coefficient defined as the ratio between the electric field $E$ on the sample plane  and the externally incident electric field $E_{\text{p}}$,
while $\alpha(\omega)$ denotes the bare transmission coefficient in  absence of the sample, see Supplemental Material (SM) Sec.~I~\cite{supp}\nocite{Cardona.1961}. 
Here we have restricted to the case that the sample is in the middle of the symmetric cavity and the  pump light is normally incident with the wave vector $k=\omega/c$.
The dimensionless factor $\beta(\omega)$ has the meaning of the radiation coefficient inside the cavity viewed from the sample.
It approaches $\beta \approx 2\pi \tan(kh/2)$ for a perfect cavity with $R \rightarrow -1$.
Therefore, both $\alpha$ and $\beta$  contain resonant poles from bare anti-node cavity photons at $kh=(2N+1)\pi$.
Through its 2D optical conductivity $\sigma(\omega)$ in \equa{eq:alpha}, the sample hybridizes with bare cavity photons and push their eigen frequencies to those of the physical cavity photons labeled by $\omega_{\text{N}}$, which will be referred to as `cavity photons' in the following.
They appear as  the poles of $\alpha^\prime$ determined by the zeros of its denominator (only anti-node cavity photons are involved). 


\emph{Step-like ponderomotive potential---}The coherent pump field drives the cavity photon which in turn drives the sample, and generates an effective static force on the slow degrees of freedom $\phi$, the ponderomotive force $F_{\text{P}}(\phi)$~\cite{sun2024floquet, Aliev:1992aa,Grimm:2000_optical_trap,Moffitt.2008,landau2013electrodynamics,RevModPhys.86.1391,wan2017control,wan2018nonequilibrium,zhou2021terahertz,zhou2023vibrational, Aliev:1992aa,  sun2018universal,Wolff.2019_P_force_graphene, Rikhter.2024, Jiang2024,braginsky1997optical,buonanno2002signal,hu2025microscopic,Wang.2025}. 
Formally, the total force on $\phi$ is $F= -\langle \partial_\phi L\rangle$ where the  bracket denotes time average for classical systems, and path-integral average over all paths of the fast fields $X$ and $A$ in the quantum mechanical case. 
By integrating out the fast degrees of freedom $X$ and $A$ using the Keldysh path integral~\cite{kamenev2023field,altland2010condensed,keldysh1964diagram,schwinger1961brownian}, one may obtain the low energy effective field theory $L[\phi]=L_0[\phi]+V_{\text{P}}[\phi]$ for $\phi$, whose potential term reads~\cite{sun2024floquet} 
\begin{eqnarray}\label{eq:FinalV}
 V[\phi]=V_0[\phi]+V_{\text{P}}[\phi],\qquad F_{\text{P}}= -\partial_\phi V_{\text{P}}
\,.
\end{eqnarray}
Here $V_0[\phi]$ is the equilibrium term that exists even at zero  pump field
and $V_{\text{P}}[\phi]$ is the ponderomotive potential induced by the pump field $A_{\text{p}}$. 
At the mean field level, the configuration  of $\phi$ that minimizes  the potential landscape $V[\phi]$ gives the non-equilibrium steady state.

In the steady state, the cavity photon field  must be oscillating periodically in time with a classically coherent value, the `mean field', together with small quantum and thermal fluctuations. Within linear response, the classical value is simply $Ae^{-i\omega t}+c.c.$ where $A=\alpha'(\omega) A_{\text{p}}$ is found from  Maxwell's equations using the optical conductivity $\sigma(\omega, \phi)$ at a fixed $\phi$. 
Therefore, the sample is now periodically driven by the cavity field $A$.
The force $F= -\langle \partial_\phi L\rangle = -\langle \partial_\phi L_{\text{M}}\rangle_X$ on $\phi$ could be computed from
$L_{\text{M}}$ in \equa{eq:initialL} with the classical cavity field  $A$ treated as the external driving field, 
which yields the ponderomotive force $F_{\text{P}}$.
The effective potential generated by the vacuum fluctuations of  $A$ is a much weaker effect and is incorporated into the $V_0[\phi]$ part of \equa{eq:FinalV}, see SM Sec.~IIC for an estimation~\cite{supp}.
In the absence of dissipation, meaning that there is no energy absorption within the sample, the driven sample belongs to case~1 below Eq.~4 of Ref.~\cite{sun2024floquet}, and the ponderomotive force must be equal to that predicted by Eq.~4 there.
At second order in the pump field, it is  simply related to the equilibrium optical conductivity $\sigma(\phi,\omega)$ of the sample as
\begin{align}\label{eqn:FP}
F_{\text{P}}= \frac{i\omega}{c^2}  |A|^2 \partial_\phi \sigma
=\frac{i\omega}{c^2}  |\alpha'|^2  A_{\text{p}}^2
\partial_\phi \sigma
\end{align}
where $\omega>0$,  $\alpha'$ is from \equa{eq:alpha} and $\sigma$   is purely imaginary in the absence of dissipation, see Appendix B for detailed derivation.

One is now ready to observe that the ponderomotive potential from \equa{eqn:FP} has a universal  step-like structure shown in \fig{fig:FPcavity}(c). 
As $\phi$ varies, the optical conductivity of the sample changes, and thus the cavity photon frequencies $\omega_{\text{N}}(\phi)$ are shifted according to the poles of $\alpha'$, as shown schematically in \fig{fig:FPcavity}(b). 
Suppose that a cavity photon $\omega_2(\phi)$ is closely above the pump frequency $\omega$ when  $\phi=\phi_0$ and red shifts
with decreasing $\phi$ as in \fig{fig:FPcavity}(b), it may hit a resonance with the pump at a `resonance' value of $\phi=\phi_{\text{c}}$. 
There the  ponderomotive force and potential from \equa{eqn:FP} are dominated by the resonant pole of $|\alpha'|^2$:
\begin{align}\label{eq:FpVp_step}
F_{\text{P}} 
\approx -\frac{V_{\text{u}}}{\pi}\frac{\gamma_\phi  }{(\phi-\phi_{\text{c}})^2+\gamma_\phi^2}
,\quad
V_{\text{P}} \approx \frac{V_{\text{u}}}{\pi}
	 \arctan \frac{\phi-\phi_{\text{c}}}{\gamma_\phi}
\end{align}
where $\gamma_\phi$ is the width of the resonance in the $\phi$ direction~\cite{1note}, see SM Sec.~II~\cite{supp}.
Therefore, the ponderomotive force has a Lorentzian peak at $\phi_c$ because the cavity photon field is resonantly enhanced here, and the ponderomotive potential has a step-like `arctan' structure, as shown in  \fig{fig:FPcavity}(c).
Amazingly, if the sample has no dissipation so that the only contribution to the linewidth of the cavity photon is its radiative loss $\gamma_{\text{r}}=\omega |T|^2$ due to a nonzero transmission coefficient of the mirror,  the height of the potential jump is a universal value
\begin{align}\label{eq:Vu}
V_{\text{u}}= E_{\text{p}}^2 \lambda/(4\pi)
\end{align}
that depends only on the electrical field and vacuum wavelength $\lambda=2\pi/k$ of the  incident light, see SM Sec.~II~\cite{supp}.
It has the physical meaning of the 2D EM field energy density of the pump laser in a 2D layer of thickness equal to its wavelength.
We note that this is a special mechanism brought by a cavity,  
which is absent in conventional light-induced phase transitions in free-space~\cite{cavalleri2004evidence,liu2022unifying,oka2009photovoltaic,mciver2020light}.

The sign of the ponderomotive force means that it always  pushes $\phi$ in the direction so that $\mathrm{Im}[\sigma]$ becomes lower, leading to red shifts of the cavity photons. 
Note that according to the pole of $\alpha'$, if $\sigma$ itself contains no extra poles, the $N$th cavity photon mode frequency is constrained between $(2N)\pi c/h$ and $(2N+2)\pi c/h$. Therefore, at most one cavity photon could  be possibly tuned by $\phi$ to cross the resonance with the pump, meaning that there  is at most one ponderomotive  step on the potential landscape $V(\phi)$.

Together with the equilibrium potential whose structure close to the  equilibrium value $\phi_0$ is $V_0 \sim (\phi-\phi_0)^2$, the  total potential in \equa{eq:FinalV} looks like that in  \fig{fig:FPcavity}(d).
Because of the step-like ponderomotive potential, a dip appears closely to the left of $\phi_{\text{c}}$, creating a new potential minimum.
When the pump strength of $E_{\text{p}}$ exceeds a critical value $E_{\text{c}}$, the new minimum becomes lower than the original one so that a first-order phase transition from $\phi=\phi_0$ to $\phi=\phi_{\text{c}}$ occurs. 
Note that the cavity photon degree of freedom plays an essential role in inducing this phase transition by endowing the ponderomotive potential with the step-structure in \equa{eq:FpVp_step}.

%
%

In the following, we apply this theory to two systems in driven cavities, 
and demonstrate the first order phase transitions induced by the ponderomotive step potential in \equa{eq:FpVp_step}.

\begin{figure}
    \includegraphics[width=\linewidth]{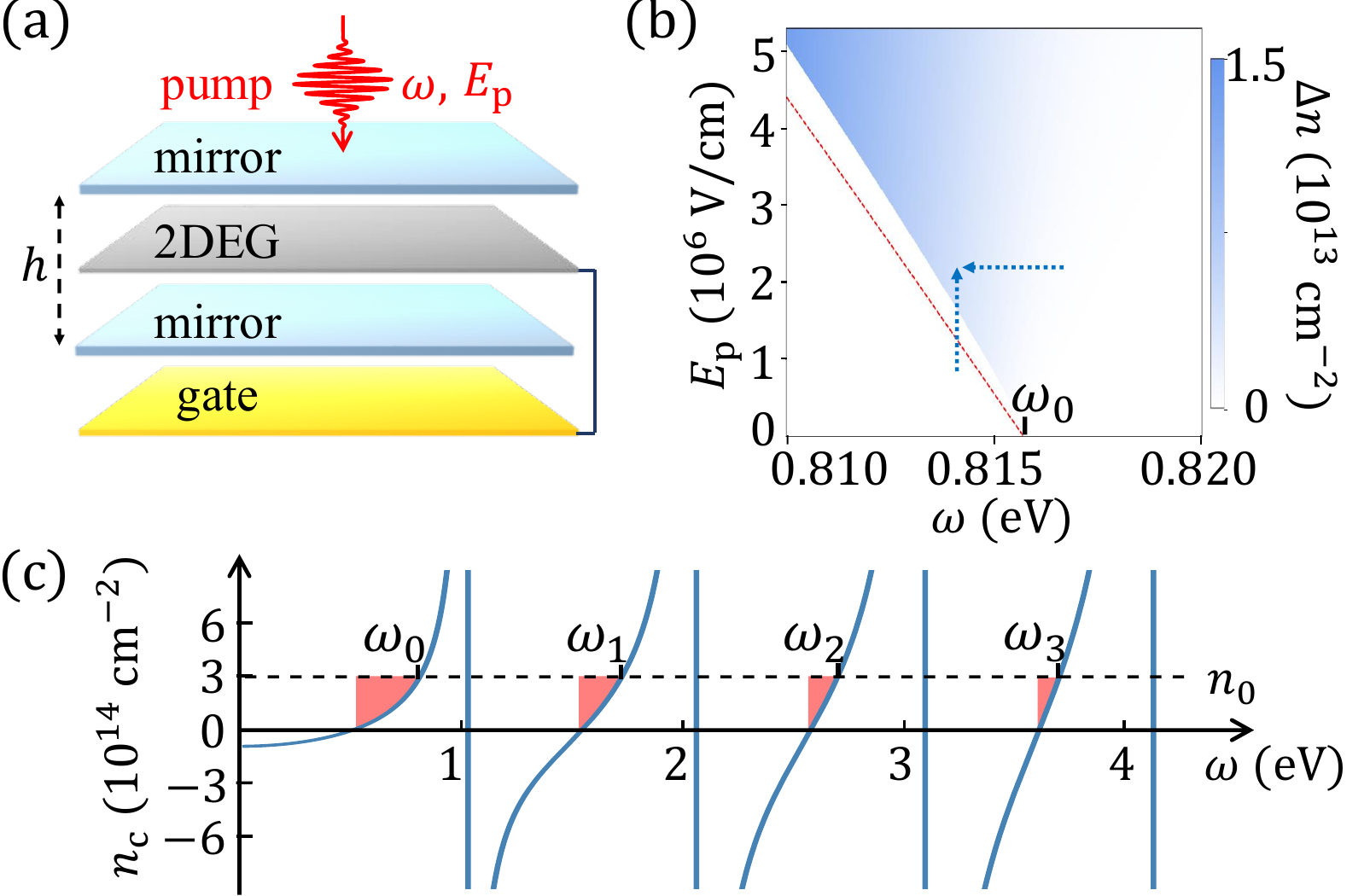}
    \caption{ (a) Schematic of the 2D electron gas in the driven cavity. The 2DEG is electrically connected to a gate outside of the cavity. 
    (b) Phase diagram of the device on the plane of pump frequency $\omega$ and electric field $E_{\text{P}}$, where the color scale corresponds to the calculated $\Delta n$, showing a clear phase boundary. The red dashed line shows the analytical prediction for the phase boundary from \equa{eq:Ec} with the damping correction.
    The parameters are $m=0.01 \unit{m_e}$, $n_0=3\times10^{14} \unit{\mathrm{cm}^{-2}}$, $h=1.2 \unit{\mu m}$, $C=0.1 \unit{\mu \mathrm{F}/\mathrm{cm}^{-2}}$, $T^2=1-R^2= 2 \times 10^{-4}$, and $\gamma = 1 \unit{meV}$ for the Drude optical conductivity. 
	(c) The blue curve is $n_c$ as a function of $\omega$. The black dashed line represents $n_0$ and the red zone highlights its difference from the possible new phase. 
    \label{fig:ScheGas}}
\end{figure}

\emph{A 2DEG in a driven cavity~\cite{zhang_collective_2016}} is the first example, as shown in \fig{fig:ScheGas}(a). The 2DEG is electrically connected to a gate electrode outside of the cavity, so that its carrier density is allowed to change. 
It is described by the Lagrangian in \equa{eq:initialL} with the matter part being
\begin{align}\label{eq:eGasL}
    L_{\mathrm{M}} 
    &=\int d^2r  \bar{\psi} 
    \left[
    -i\partial_t + \xi(\hat{\mathbf{p}}+\mathbf{A})
    \right] \psi 
    + V_{\text{i}}
    +V_{\text{ee}}
\end{align}
where $\bar{\psi}$ and $\psi$ are the creation and annihilation fermion fields for electrons containing the fast degrees of freedom, while the slow degree of freedom $\phi$ considered here is the uniform electron density $n=\langle \bar{\psi}\psi \rangle $ on the sample,
and $\hat{\mathbf{p}}=-i\nabla$ acts on the fermion field. 
The  $V_{\text{i}}$  term represents electron-impurity scattering,  and $V_{\text{ee}}$ is the electron-electron interaction including with those in the gate.
For notational simplicity, the Planck constant $\hbar$, the elementary charge $e$ and the speed of light $c$  are set to be 1 but will be restored when necessary.

For a parabolic band $\xi(\mathbf{p})= p^2/(2m)-\mu$ without impurity scattering, there is a quick derivation of the ponderomotive force because the paramagnetic coupling between electrons and light is inconsequential, and the only relevant coupling is the diamagnetic term: $nA^2/(2m)$.
Taking a derivative with respect to $n$, the ponderomotive force for the electron density is obtained as
\begin{align}\label{eq:Fp_n}
    F_{\text{P}} &=-\frac{e^2}{2m} \langle A(t)^2 \rangle =-\frac{e^2}{m}|\alpha'|^2A_{\text{p}}^2
\end{align}
in agreement with \equa{eqn:FP},
where  $\alpha'$ is from \equa{eq:alpha}. Indeed, this is what we have proved in \equa{eqn:FP} considering that 
the  optical conductivity  $\sigma = i ne^2/(m\omega)$  is proportional to $n$.
The electron-electron interaction does not modify this result owing to the Galilean invariance of a parabolic-band electron liquid in the clean limit~\cite{mah00, Chubukov.2012, Chubukov.2012_2, Sun.2018_third}.
This  ponderomotive force  pushes the carrier density to smaller values so that the cavity photon frequency shifts down. 
Therefore, the ponderomotive potential felt by $n$ is simply the step-like potential in \equa{eq:FpVp_step} and \fig{fig:FPcavity}(c)  with $\phi$, $\phi_{\text{c}}$ and $\gamma_{\phi}$ replaced by $n$, $n_{\text{c}}$ and $\gamma_{\text{n}}$ whose values are found from the poles of $\alpha'$ in \equa{eq:alpha}. 
The total potential is \equa{eq:FinalV} with the equilibrium term being
$
V_0(n)= e^2(n-n_0)^2/(2C)
$
where $n_0$ is the equilibrium carrier density, and the charge stiffness  $1/C=1/C_0+1/\nu$ is set by the capacitance $C_0=1/(4\pi d)$ between the sample and the gate and the electronic density of states $\nu$ near the fermi surface. 
Therefore,  the potential landscape is \fig{fig:FPcavity}(d) with  $\phi$ being the carrier density $n$.

When the carrier density $n$ is shifted to the `resonant density' set by the poles of $\alpha'$ in \equa{eq:alpha}:
\begin{align}\label{eqn:nc}
n_{\text{c}}=-\frac{c \omega  m}{e^2 \beta}
\approx
-\frac{c \omega  m}{2\pi e^2 \tan(kh/2)}
\,,
\end{align}
a resonance between a cavity photon mode  and the pump   occurs and $A$ and $F_{\text{P}}$ reach their maxima as shown in \fig{fig:FPcavity}(c).
The resonant density $n_c$ from \equa{eqn:nc}  is plotted as a function of pump frequency in \fig{fig:ScheGas}(c). 
Note that $n_{\text{c}}(\omega)$ has
periodic `divergences' around $kh=2N \pi$ for $N=1,2,3, ...$, and
 periodic  zeros at $kh=(2N-1)\pi$ when the pump is at resonances with  bare  anti-node cavity  photon modes. 
From \fig{fig:FPcavity}(d), it is obvious that if  $0<n_{\text{c}}<n_0$ (red region in \fig{fig:ScheGas}(c)), the step-like drop of potential creates a new potential minimum at $n_{\text{s}} \lessapprox n_{\text{c}}$. 
Physically, $n_{\text{c}}<n_0$ means that there is a  cavity photon mode moderately above the pump frequency, so that if the carrier density is pushed down, the photon red shifts and potentially hits a resonance with the pump, see \fig{fig:FPcavity}(b).

Due to the sharpness of the step potential, the energy difference between the two minima is  simplified to $V(n_{\text{s}})-V(n_0) \approx V_0(n_{\text{c}})-V_{\text{u}}$. 
If the external pump field is stronger than the critical value $E_{\text{c}}$ set by 
$V_{\text{u}}(E_{\text{c}})=V_0(n_{\text{c}})$, i.e., 
\begin{align}\label{eq:Ec}
E_{\text{c}} & \approx e\sqrt{\frac{k}{C}}(n_0-n_{\text{c}})
\end{align}
the energy of the new minimum becomes lower than the original one, and the system undergoes a first order phase transition to this state of lower carrier density. 
This result may be relevant to the transitions observed numerically in
driven electron-phonon systems~\cite{yarmohammadi2023nonequilibrium,yarmohammadi2024ultrafastdynamicsfermionchain}.
Note that if $n_{\text{c}}>n_0$, the drop of ponderomotive potential occurs to the right of $n_0$ in \fig{fig:FPcavity}(d), and would not create a new minimum.




A numerically exact phase diagram computed from minimizing \equa{eq:FinalV} is shown in Fig.~\ref{fig:ScheGas}(b) where the color scale represents the (negative) change of carrier density $\Delta n = n_0 - n_{\text{s}}$ of the driven state compared to the equilibrium state. 
As the pump field  $E_{\text{p}}$ exceeds the phase boundary following the vertical arrow, the system undergoes a first-order phase transition to the new phase where the  electron density shifts abruptly from $n_0$ to $n_{\text{s}}$ so that its relevant photon mode becomes near resonance with (but slightly lower than) the pump. 
If one follows the horizontal arrow in Fig.~\ref{fig:ScheGas}(b), no phase transition occurs. At the initial point where $n_{\text{c}}>n_0$,  there is only one minimum of $V(n)$ located at $n_{\text{s}} \lessapprox n_0$. As the pump frequency decreases so that $n_{\text{c}}$ decreases to below $n_0$,  the global minimum $n_{\text{s}}$   sticks  to the left of $n_{\text{c}}$ and moves smoothly with it, see \fig{fig:FPcavity}(d). 
This type of phase diagram resembles that of the liquid-gas transition whose phase boundary terminates at a critical point.

Note that  for pumping frequencies closely below the equilibrium cavity photon modes labeled by $\omega_\text{N}$ in Fig.~\ref{fig:ScheGas}(c), the critical pump field $E_{\text{c}}$ actually approaches zero because it only takes a small amount of carrier density shift for the photon to red shift to resonate with the pump. Therefore, these are the best frequencies to achieve this new phase  experimentally.
To avoid over heating in an actual experiment, an ultrafast laser pulse may be applied with its parameters tuned in time following the horizontal and vertical arrows in Fig.~\ref{fig:ScheGas}(b). 
The former steers the system to the new minimum which manifests as a dip/peak of the reflected/transmitted photons, while the later does not because there is not enough time to overcome the energy barrier.

When there is a nonzero scattering rate $\gamma$ so that the optical conductivity becomes the Drude form $\sigma = ine^2/[m(\omega+i\gamma)]$, it turns out that Eqs.~\eqref{eq:Fp_n}\eqref{eq:FpVp_step} still hold, while the  potential drop $V_{\text{u}}$ in \equa{eq:Vu} is suppressed by a factor of about $\gamma_{\text{r}}/(\gamma_{\text{r}}-2\gamma \sin kh)$ where $\sin kh<0$ in the phase transition region (see SM Sec.~III~\cite{supp}).
The critical field is thus larger than \equa{eq:Ec} by the inverse square root of this factor.

\emph{Dirty superconductors---}A more intriguing example is a BCS-type dirty superconductor, either a thin film or a 2D superconductor, in a driven THz cavity~\cite{zhang_collective_2016, krupka2005high, cuper2024conductivity} with a configuration shown in \fig{fig:FPcavity}(a).
In this device, the sample is not electrically connected to a gate so that its charge density is not allowed to change. Instead, the order parameter $\Delta$ for the superconducting state can vary, which is the low energy degree of freedom of the system.  The matter part of the Lagrangian density in \equa{eq:initialL}  could be written as~\cite{Sun.2020_superconductor}
\begin{align}
  \mathcal{L}_{\text{M}}=\bar{\Psi}&
    \begin{pmatrix}
        -i\partial_t+\xi(\hat{\mathbf{p}}+\mathbf{A})
         & \Delta 
\\
        \Delta^{\ast} & -i\partial_t-\xi(-\hat{\mathbf{p}}+\mathbf{A})
    \end{pmatrix}\Psi
    \nonumber\\
    +& V_{\mathrm{dis}}(\mathbf{r}) n(\mathbf{r})
    + \frac{1}{g}|\Delta|^2
    ,
\label{eq:LforSC}
\end{align}
where $\Psi=(\psi_\uparrow, \bar{\psi}_\downarrow)^T$ is the fermion field for spin up and down,
$\xi(\mathbf{p})=\varepsilon(p)-\mu$ is the electronic kinetic energy same as \equa{eq:eGasL},
$V_{\mathrm{dis}}$ represents the disorder potential, and $g$ is the s-wave attractive interacting strength between electrons.

\begin{figure}
	\includegraphics[width=\linewidth]{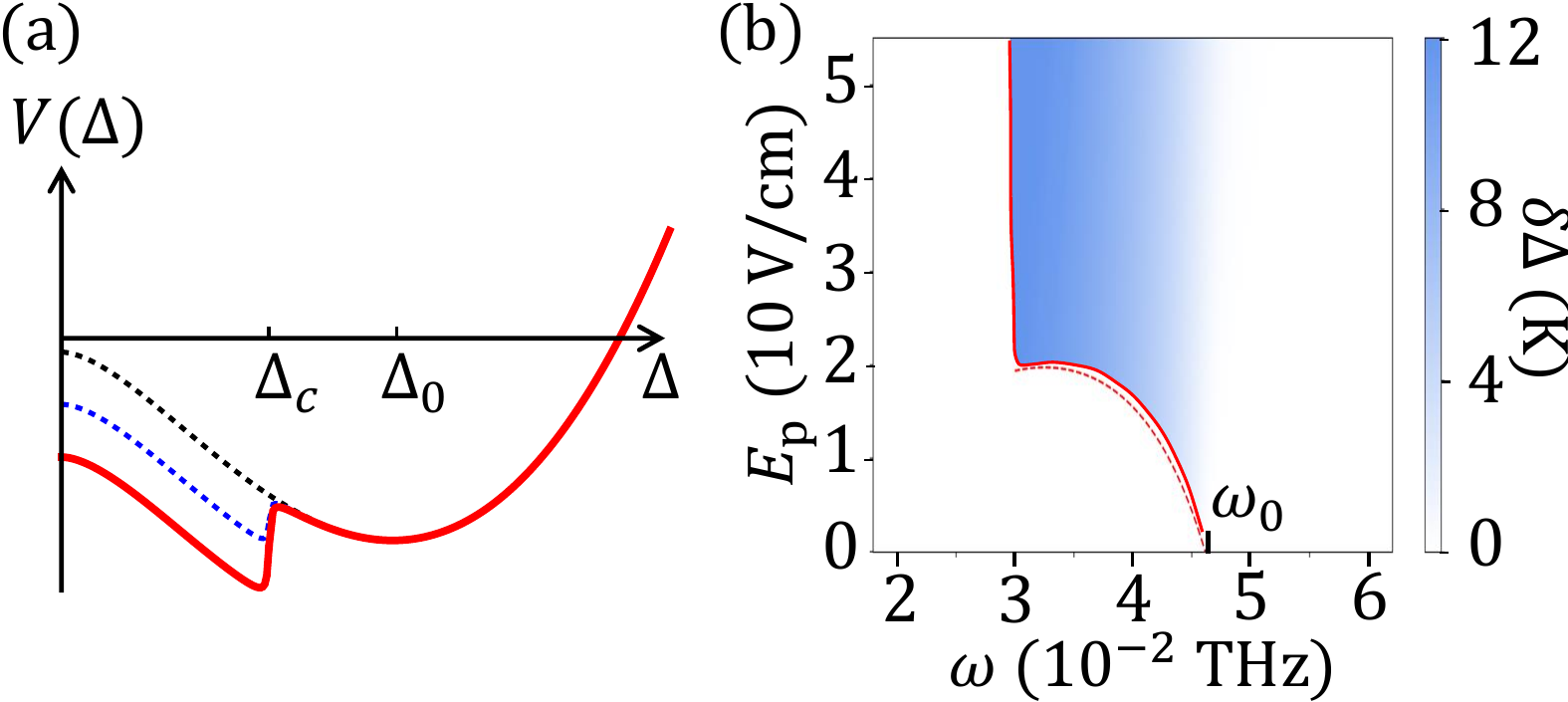}
	\caption{\label{fig:ScheSC} (a) The total potential felt by the order parameter $\Delta$ of a superconductor in the device in \fig{fig:FPcavity}(a) for $E_{\text{p}}=0$ (black dashed line), $E_{\text{p}}\approx E_{\text{c}}$ (blue dashed line), and $E_{\text{p}}>E_{\text{c}}$ (red  solid line). (b) Numerically exact phase diagram of the device where the color scale  represents $\delta\Delta = \Delta_0-\Delta_{\text{s}}$, the (negative) change of gap  relative to its equilibrium value. 
	A clear phase boundary is marked by the red solid curve, while the red dashed line is the analytical approximation to it.
		The parameters are $m=0.3 \unit{m_e}$, $\tau=0.03 \unit{ps}$, $\Delta_0=12\unit{K}$, $g\nu=0.5$, $n_0=1\times10^{13}\mathrm{cm}^{-2}$, $h=5\unit{mm}$ and $R =0.99$. 
	}
\end{figure}

We now study the ponderomotive force exerted by the pump $A_{\text{p}}$ on the order parameter $\Delta$ which is taken to be real without loss of generality.
To capture the essential physics without introducing much complexity, we discuss the case of zero temperature with the equilibrium gap $\Delta_0$ so that there are no thermally excited quasi-particles.
For a dirty superconductor, the Mattis-Bardeen theory~\cite{mattis1958theory} renders a good approximation to the optical conductivity, which at low frequencies ($\omega\ll 2\Delta$) is just the superfluid response
\begin{align}\label{eqn:sigma_SC}
\sigma(\omega)\approx i\frac{ne^2}{m}\frac{\pi \tau \Delta}{\omega}
\,.
\end{align}
Here $n$ is the total carrier density, $\tau$ is the disorder limited electronic mean free time, and the superfluid density $n \pi \tau \Delta$ is proportional to the gap. 
Therefore, similar to the case of a 2DEG, the oscillating electric field tends to push the gap to smaller values to reduce the superfluid density, so that the cavity photon gets red-shifted.

Formally, for a sub-gap pump $\omega<2\Delta_0$, there is no dissipation from quasiparticles or pair-breaking excitation. 
Therefore, \equa{eqn:FP} and \equa{eq:FpVp_step} apply here. The ponderomotive force felt by $\Delta$ is simply
$
F_{\text{P}}=-\pi n\tau \frac{e^2}{m}A^2=-\pi n\tau \frac{e^2}{m}|\alpha^\prime|^2 A_{\text{p}}^2 
$,
and the ponderomotive potential $V_{\text{P}}$  is  the step-like potential in \equa{eq:FpVp_step}  with $\phi$, $\phi_c$ and $\gamma_{\phi}$ replaced by $\Delta$, $\Delta_{\text{c}}$ and $\gamma_{\Delta}$ whose values are found from the poles of $\alpha'$.
The `resonance gap'
\begin{eqnarray}
	\Delta_{\text{c}}=-\frac{1}{\pi n\tau} \frac{c \omega  m}{e^2 \beta}
	\label{eq:SCDc}
\end{eqnarray}
simply has an extra constant factor compared to $n_{\text{c}}$ in \equa{eqn:nc} so that its frequency dependence looks like \fig{fig:ScheGas}(c). 
When the order parameter is shifted to $\Delta_{\text{c}}$, the cavity photon frequency is shifted to a point in resonance with the pump, and a step-like drop of $V_{\text{P}}$ occurs.

%



The equilibrium potential $V_0(\Delta)= 	\Delta^2/g-\nu\Delta^2\ln(\Lambda/\Delta)$ is  from standard BCS theory~\cite{Sun.2020_superconductor, anderson1959theory} with $\nu$ being the density of states and $\Lambda$ being the energy cut-off, whose minimum gives the equilibrium gap $\Delta_0=\Lambda e^{-1/(g\nu)-1/2}$.
With the pump, the total potential $V(\Delta)=V_0+V_{\text{P}}$ is modified to that in \fig{fig:ScheSC}(a). 
If  $0<\Delta_{\text{c}}<\Delta_0$, the step-like drop of potential occurs to the left of $\Delta_0$ as shown  in \fig{fig:ScheSC}(a), creating a new energy minimum at $\Delta_{\text{s}}$.
A strong enough pump would make the energy of the new minimum lower than the original one, inducing a first order phase transition of the superconducting order parameter from $\approx\Delta_0$ to  $\Delta_{\text{s}} \lessapprox \Delta_{\text{c}}$.
Similar to \fig{fig:ScheGas}(b), the numerically exact phase diagram is shown in Fig.~\ref{fig:ScheSC}(b) which displays a clear phase boundary. 

Since the superconductor introduces no extra loss to the cavity photon, the step-like drop of the ponderomotive potential  is the same universal value 
$
V_{\text{u}} =  E_{\text{p}}^2 /(2k)
$.
The critical field $E_{\text{c}}$ could be estimated simply by $V_0(\Delta_{\text{c}})-V_0(\Delta_0)=V_{\text{u}}(E_{\text{c}})$ which gives the red dashed curve in Fig.~\ref{fig:ScheSC}(b), a decent approximation to the phase boundary.
If $\Delta_{\text{c}}$ is close to $\Delta_0$ which happens for pump frequencies closely below the equilibrium cavity photons, the equilibrium potential  could be approximated by a quadratic one so that the criterion becomes $\nu(\Delta_{\text{c}}-\Delta_0)^2=V_{\text{u}}(E_{\text{c}})$ 
and renders a very small critical field $E_{\text{c}}=(\Delta_0-\Delta_{\text{c}})\sqrt{2k \nu}$~\cite{3note}. 

%

To conclude, we note that Eqs.~\eqref{eqn:FP} \eqref{eq:FpVp_step} for the dirty superconductor is only strictly correct at zero temperature so that there is no optical absorption from quasi-particles, 
while the Eliashberg effect~\cite{PismaZhETF.11.186,klapwijk1977radiation, curtis2019cavity} that enhances superconductivity dominates at temperatures close to the critical temperature,
see Appendix~C for nonzero temperature effects.
Although we focused on Fabry–Pérot cavities for analytical clarity, the other types of cavities such as plasmonic/polaritonic nano cavities can also be treated within our framework with the coefficient $\alpha^\prime(\omega, \phi)$ replaced by the appropriate expressions.
Future directions include  exploring other new phases induced by this potential in driven cavities, the effect of fluctuations beyond the mean-field level, and the force induced by vacuum fluctuations in equilibrium cavities~\cite{felice2022breakdown,Rubio.2023,Faist.2024, enkner2024enhancedfractionalquantumhall, yang2025quantumhalleffectchiral, keren2025cavity}.

\begin{acknowledgments}
	This work is supported by the National Key	Research and Development Program of China (2022YFA1204700),  the National Natural Science Foundation of China (Grants No. 12374291 and No. 12421004), Beijing Natural Science Foundation (Z240005),  and the startup grant from Tsinghua University. 
	We thank M. Yarmohammadi, Q. Xiong, T. Qin and T. Xiao for helpful discussions.
\end{acknowledgments} 

\emph{Data availability—}Numerical codes and data for plots in this paper are available online~\cite{codesanddata}.




\begin{thebibliography}{88}
\bibitem [{3no()}]{3note}%
  \BibitemOpen
  \href@noop {} {}\bibinfo {note} {In the other limit of $\Delta_{\text{c}} \ll \Delta_0$, the criterion becomes $-V_0(\Delta_0)=V_{\text{u}}(E_{\text{c}})$ and the critical field is $E_{\text{c}} \approx \Delta_0\sqrt{\nu k}$. However, note that if $\Delta_{\text{c}}$ is so small that $2\Delta_{\text{c}} < \omega$, the pump would cause pair breaking excitations and \equa{eq:FpVp_step} no longer holds.}\BibitemShut {Stop}%
\end{thebibliography}
%

\onecolumngrid
\begin{center}
\textbf{\large{{End Matter}}}
\end{center}
\twocolumngrid

\appendix	
\setcounter{equation}{0}
\renewcommand{\theequation}{A\arabic{equation}}
\emph{Appendix A: Heuristic derivation of the step-like ponderomotive potential---}The step-like structure of the pondermotive potential could be heuristically understood using a linearly driven harmonic oscillator~\cite{sun2024floquet}:
\begin{align}\label{eqn:oscillator}
L= \frac{1}{2} \left[ -\dot{X}^2+(\omega_{0}^2+\phi) X^2 \right]   + E(t)X
\,
\end{align}
where $X$ is the displacement of the oscillator and $E(t)=E_0 e^{-i\omega t} +c.c.$ is the driving field.
The slow degree of freedom $\phi$ couple to the oscillator by shifting its intrinsic frequency.
Following the derivation of Eq.~11 in Ref.~\cite{sun2024floquet}, the  force experienced by $\phi$ is simply 
$F=-\partial_\phi L=-\frac{1}{2} X^2 $. 
Taking the driving-field-dependent part of its time average, one obtains the  ponderomotive force $F_{\text{P}}=-|\chi_R(\omega) E_0|^2$ where $\chi_R(\omega)=1/(-\omega^2-i\gamma \omega + \omega_0^2+\phi)$ is the retarded response function of the oscillator.
We have added a phenomelogical damping rate $\gamma$, which 
could be formally included in the Keldysh action version of \equa{eqn:oscillator}~\cite{sun2024floquet}.
Therefore, one uncovers the Lorentzian structure of the  ponderomotive force  and the step-like structure of the ponderomotive potential: 
\begin{align}\label{eq:Fp_simple_SI}
F_{\text{P}}= -\frac{E_0^2}{(\phi-\phi_c)^2+\gamma_\phi^2}
,\,\,
V_{\text{P}}=\frac{E_0^2}{\gamma_\phi} \arctan\left[\frac{\phi-\phi_c}{\gamma_\phi} \right]
\end{align}
where $\phi_c=\omega^2- \omega_0^2$ and $\gamma_\phi=\gamma \omega $.
Although derived from the classical equation of motion, this result is exact because of the exact quantum-classical correspondence of the retarded response of a harmonic oscillator.
This force reflects a generic phenomenon of the ponderomotive force on a slow degree of freedom $\phi$ that affects a driven harmonic oscillator by shifting its  eigen frequency:  it points in the direction that red shifts the oscillator and results in a step-like potential.

\setcounter{equation}{0}
\renewcommand{\theequation}{B\arabic{equation}}
\emph{Appendix B: Ponderomotive force in an optically driven cavity---}
The generic Lagrangian of the system is \equa{eq:initialL} 
with the electromagnetic part being $L_{\mathrm{EM}}=L_{\mathrm{cavity}}[A]+L_{\mathrm{pump}}[A,A_{\text{p}}]$.
Here 
$L_{\mathrm{pump}}
=\int d\mathbf{r} d\mathbf{r}'A(\mathbf{r})T_{\mathbf{r}\mathbf{r}'}
A_{\text{p}}(\mathbf{r}')
$ 
describes  the linear coupling between the cavity photonic degree of freedom $A$ and the pump $A_{\text{p}}$ with the kernel $T_{\mathbf{r}\mathbf{r}'}$.  
We will show that our results can be derived in terms of the transmission and reflection coefficients $R(\omega)$/$T(\omega)$ of the mirror without knowing the explicit forms of $T_{\mathbf{r}\mathbf{r}'}$.

The Keldysh path integral description of the system is
\begin{align}\label{GeneratingFunctional}
  Z&=\int D[X_{\text{cl/q}},\phi_{\text{cl/q}},A_{\text{cl/q}}]e^{-iS[X_{\text{cl/q}},\phi_{\text{cl/q}},A_{\text{cl/q}};A_{\text{p}}]}
  \notag\\
  &=\int D[\phi_{\text{cl/q}}]e^{-iS_\phi[\phi_{\text{cl/q}};A_{\text{p}}]}
\end{align}
defined on the closed time contour~\cite{kamenev2023field,altland2010condensed,keldysh1964diagram,schwinger1961brownian}. 
Here the subscripts `cl' and `q' denote the `classical' and `quantum' components of the fields. 
After integrating out the fast degrees of freedom $X$ and $A$, one is left with the effective low energy theory for $\phi$ with the Keldysh action:
\begin{align}\label{Keldysh_phi}
S_\phi[\phi_{\text{cl/q}}]
&=-\sum_\omega
\left[F_0(\phi_{\text{cl}})+F_{\text{P}}(\phi_{\text{cl}},A_{\text{p}})\right]_{-\omega}
\phi_{\text{q}}(\omega)
+
O(\phi_{\text{q}}^2)
,\notag\\
F_0+F_{\text{P}}
& =\langle -\partial_{\phi_{\text{q}}}S \rangle_{X_{\text{cl/q}},A_{\text{cl/q}}}
|_{\phi_{\text{q}}=0}
\,.
\end{align}
Here the zero frequency limit of $F_0+F_{\text{P}}$ is the static force that $\phi$ feels, which is formally the Keldysh path integral average of the `force operator'.
Its driving-field induced part $F_{\text{P}}(\phi_{\text{cl}},A_{\text{p}})$ is defined as the ponderomotive force~\cite{sun2024floquet}.

We now show that under the `no dissipation' condition for the material, the ponderomotive force $F_{\text{P}}$ on $\phi$ could be directly inferred from the linear and nonlinear optical response functions of the material.
The `no dissipation' condition  means that there is no extra bath for the \emph{material} and no resonant transitions in it. However, the cavity photon itself can have dissipation which is implied by nonzero transmission coefficient of the mirrors. 
Under this condition, the steady state exists, and the fields $X$ and $\phi$ on the 
forward (denoted by `$+$') 
and backward (`$-$')  time contours are decoupled, meaning the action could be written as~\cite{sun2024floquet}:
\begin{align}\label{actionAppendix}
    S = S_+-S_-+S_{\text{EM}}[A_{\text{cl}},A_{\text{q}};A_{\text{p}}]
\end{align}
where $S_{\pm} = \int dtL_{\text{M}}[X_\pm,\phi_\pm,A_\pm]$.
Therefore, the two path integrals over $X_\pm$ decouple from each other, after which one obtains $S_{\pm} = \sum_{n=1}^\infty \chi^{(2n-1)}(\phi_\pm)A_\pm^{2n}$ for constant $\phi$. 
The total action  is therefore:
\begin{align}\label{S_phi_A}
    S=&\sum_{n=0}^\infty \left[\phi_{\text{q}}\partial_{\phi_{\text{cl}}}\chi^{(2n-1)}(\phi_{\text{cl}})A_{\text{cl}}^{2n}+ 2n\chi^{(2n-1)}(\phi_{\text{cl}})A_{\text{q}}A_{\text{cl}}^{2n-1} \right]\nonumber\\
    &+O(\phi_{\text{q}}A_{\text{q}}^2, \phi_{\text{q}}^3, A_{\text{q}}^3)+S_{\text{EM}}[A_{\text{cl}},A_{\text{q}};A_{\text{p}}].
\end{align}
Because the coefficients of the $O(A_{\text{q}})$ terms should be the retarded current response functions to $A$, one obtains the relation $\chi^{(2n-1)}(\phi_{\text{cl}})=-\chi^{(2n-1)}_{\text{R}}(\phi_{\text{cl}})$ after absorbing the factors into its definition. 
Note that for notational simplicity, we have suppressed the frequency arguments of the fields and the response functions. 
Integrating out the cavity photon $A$ renders the ponderomotive force 
\begin{align}\label{PforceAppendix}
	F_{\text{P}}=\sum_{n=1}^\infty
	\left[ \partial_{\phi_{\text{cl}}}\chi_R^{(2n-1)}(\phi_{\text{cl}})
	\right]
	\langle A_{\text{cl}}^{2n}
	+O(A_q^2)
	\rangle_{A_{\text{cl/q}}}|_{\phi_{\text{q}}=0}
	\,
\end{align}
on $\phi$ by comparing to \equa{Keldysh_phi}.
The $\langle  ... \rangle$ term means the expectation value of $A_{\text{cl}}^{2n}
+O(A_q^2)$ on the material that is
induced by the pump $A_{\text{p}}$ taking into account electromagnetic response  of the material, as indicated by \equa{S_phi_A} in the $\phi_q=0$ limit. 
At second order in $A_{\text{p}}$, only the linear response property of the material is concerned, which is related to its optical conducitivity $\sigma(\omega,\phi)$ as $\chi_R^{(1)}=i\omega \sigma/c^2$.
As a result, plugging the classical solution $A_{\text{cl}}=\alpha'(\omega,\phi)A_{\text{p}}$ (see \equa{eq:alpha} and SM Sec.~I~\cite{supp}  for the derivation) into \equa{PforceAppendix} 
renders  \equa{eqn:FP}, which is the exact  ponderomotive force to second order in the pump field.



\setcounter{equation}{0}
\renewcommand{\theequation}{C\arabic{equation}}
\emph{Appendix C: Light induced ponderomotive force in a dirty superconductor---}In this section, we calculate the light induced ponderomotive force on the gap of a dirty superconductor at nonzero temperatures. 
We will show that as the temperature $T$ gets close to the critical temperature $T_{\text{c}}$, the ponderomotive force may change its sign to enhance superconductivity, recovering the Eliashberg effect~\cite{PismaZhETF.11.186,klapwijk1977radiation,curtis2019cavity}.

To proceed, we employ a  simplified Hamiltonian for a dirty superconductor as an approximation to \equa{eq:LforSC}:
\begin{align}\label{eq:H1forSC}
  H=\sum_k\begin{pmatrix}
        c^\dagger_{k\uparrow}
         & c_{-k\downarrow}
    \end{pmatrix}
    \begin{pmatrix}
        \xi_k
         & \Delta 
\\
        \Delta^\ast & -\xi_k
    \end{pmatrix}
	\begin{pmatrix}
        c_{k\uparrow}
\\
        c^\dagger_{-k\downarrow}
    \end{pmatrix}
+ \frac{1}{g}|\Delta|^2
    \nonumber\\
    + MA\sum_{kk^\prime}\begin{pmatrix}
        c^\dagger_{k\uparrow}
         & c_{-k\downarrow}
    \end{pmatrix}
	\begin{pmatrix}
        c_{k'\uparrow}
\\
        c^\dagger_{-k'\downarrow}
	\end{pmatrix}
+\frac{nA^2}{2m}
\,.
\end{align}
The subscript $k$ is not the momentum but rather the index of the exact single-electron eigen states created by $c^\dagger_{k,s}$ in the disorder potential $V_{\mathrm{dis}}(\mathbf{r})$, while $-k$ labels its time-reversal counterpart. 
The electromagnetic vector potential $A$ couples electrons between these states with a constant matrix element $M$ without the restriction of `momentum' conservation, which is the essential ingredient of Mattis--Bardeen theory. 
The last term in \equa{eq:H1forSC} is the diamagnetic term with $n$ being the total carrier density.
The linear current response from \equa{eq:H1forSC} gives the Mattis--Bardeen  optical conductivity $\sigma$~\cite{mattis1958theory,tinkham2004introduction}, which fixes the parameter $M$. We take $\Delta$ to be a positive real number and compute the force on its amplitude direction without loss of generality. 

\begin{figure}
	\includegraphics[width=\linewidth]{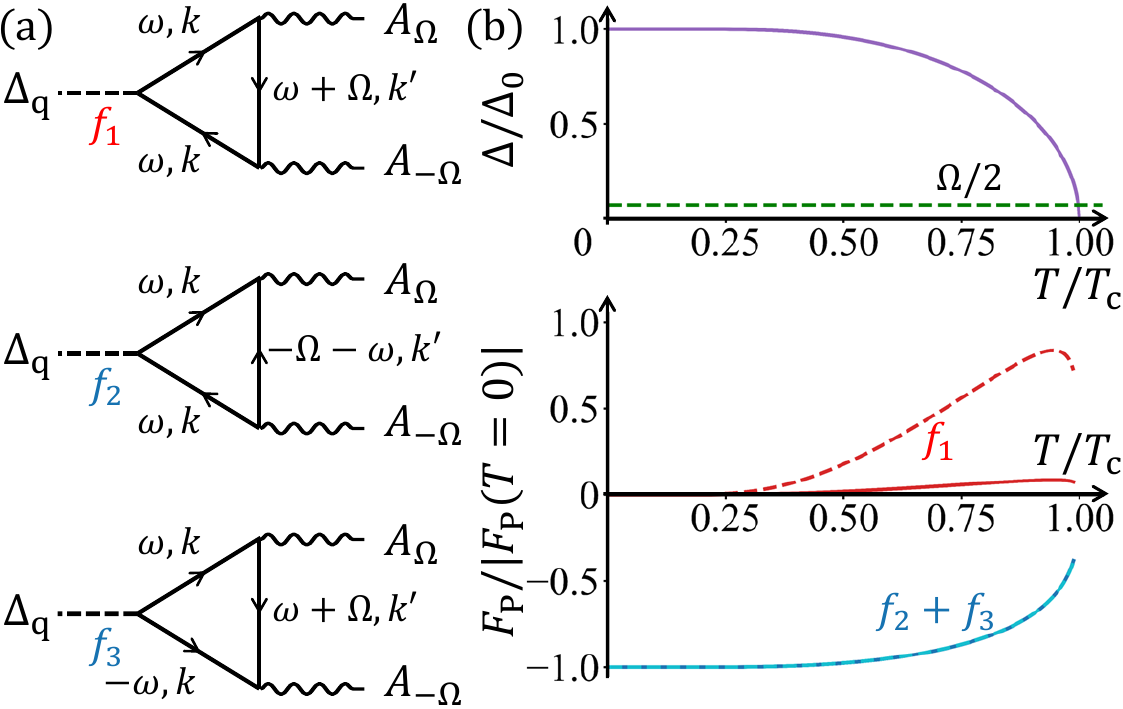} 
    \caption{
    (a) The three Feynman diagrams contributing to the ponderomotive force $F_{\text{P}}$ at second order in $A$. 
    Solid lines  are the $2\times2$ Green functions of quasi-particles in Keldysh notation with their arrows pointing from particle creation to annihilation.
    (b) The top panel shows the equilibrium superconducting gap as a function of temperature (blue curve). The dashed line shows half of the driving frequency. 
    The bottom panel shows the $f_1$ (red curves) and $f_2+f_3$ (blue/cyan curves) components of the ponderomotive force normalized by $|F_{\text{P}}|$ at zero temperature.
    The solid lines correspond to $\eta/\Delta_0=1/10$ and dashed lines correspond to $\eta/\Delta_0=1/100$. 
    The parameters are	$g\nu =0.5$, $\Delta_0 = 12\unit{K}$ and $\Omega = 1.7 \unit{K}$ ($0.035\unit{THz}$).
    }
	\label{fig:DisorderS} 
\end{figure}

In terms of the Bogoliubov quasi-particles $\left(\begin{smallmatrix}\alpha_{k\uparrow}\\\alpha^\dagger_{-k\downarrow}\end{smallmatrix}\right)=\left(\begin{smallmatrix}\cos \theta_k&\sin \theta_k\\\sin \theta_k&-\cos \theta_k\end{smallmatrix}\right)
\left(\begin{smallmatrix}c_{k\uparrow}\\c^\dagger_{-k\downarrow}\end{smallmatrix}\right)$
where $\cos \theta_k=\frac{\xi_k}{E_k}$, $\sin \theta_k=\frac{\Delta}{E_k}$ and $E_k=\sqrt{\xi_k^2+\Delta^2}$, the Hamiltonian reads

\begin{align}\label{eq:H2forSC}
  H=\sum_{k}\begin{pmatrix}
        \alpha^\dagger_{k\uparrow}
         & \alpha_{-k\downarrow}
    \end{pmatrix}
    \begin{pmatrix}
        E_k
         & 0
\\
        0 & -E_k
    \end{pmatrix}
	\begin{pmatrix}
        \alpha_{k\uparrow}
\\
        \alpha^\dagger_{-k\downarrow}
    \end{pmatrix}
  + \frac{1}{g}\Delta^2
\nonumber\\
    + MA\sum_{kk^\prime}\begin{pmatrix}
        \alpha^\dagger_{k\uparrow}
         & \alpha_{-k\downarrow}
    \end{pmatrix}\hat{W}_{kk^\prime}
	\begin{pmatrix}
        \alpha_{k'\uparrow}
\\
        \alpha^\dagger_{-k'\downarrow}
	\end{pmatrix}
+\frac{nA^2}{2m},
\end{align}
where $\hat{W}_{kk^\prime}=\left(\begin{smallmatrix}\cos (\theta_k-\theta_{k^\prime})&-\sin (\theta_k-\theta_{k^\prime})\\\sin (\theta_k-\theta_{k^\prime})&\cos (\theta_k-\theta_{k^\prime})\end{smallmatrix}\right)$.
From Eqs.~\eqref{Keldysh_phi} and \eqref{eq:H1forSC}, the Keldysh path integral representation of the ponderomotive force on $\Delta$ is
\begin{align}\label{EndMatterPforce}
F_{\text{P}} = -\sum_k\langle \bar{c}_{k}(\hat{\sigma}_1\otimes\gamma^{\text{q}})c_{k}\rangle
=\sum_k\langle \bar{\alpha}_{k}(\hat{V}_k\otimes\gamma^{\text{q}})\alpha_{k}\rangle,
\end{align}
where $\bar{\alpha}_{k}$ and $\alpha_{k}$ represent the four component fermion fields in the Nambu-Keldysh space~\cite{kamenev2023field},
$\hat{V}_{k}=\left(\begin{smallmatrix}-\sin 2\theta_k&\cos2\theta_k\\\cos2\theta_k&\sin 2\theta_k\end{smallmatrix}\right)$
is the kernel of the force in Nambu space
 and $\gamma^{\text{cl/q}}$ are the standard $2\times2$ matrices in Keldysh space. 
At second order in $A$, the force is computed from the Keldysh path integral as 
\begin{align}\label{EndMatterPforceResult}
F_{\text{P}} =
\nu^2M^2 |A|^2
\left(
f_1 +f_2 +f_3
\right),
\end{align}
where $f_{1/2/3}\left(\frac{\eta}{\Delta},\frac{\Omega}{\Delta},\frac{T}{\Delta}\right)$ are three dimensionless functions 
of the quasi-particle damping rate $\eta$, driving frequency $\Omega$ and temperature $T$ in units of the equilibrium gap $\Delta(T)$.
They come from the three Feynman diagrams shown in Fig.~\ref{fig:DisorderS}(a), respectively, see SM Sec.~IVB for the detailed derivation~\cite{supp}.
$f_{1/2}$ originates from the diagonal  terms of $\hat{V}_k$ in Eq.~(\ref{EndMatterPforce}), while $f_3$ arises from the off-diagonal terms. 

The numerical results are shown in Fig.~\ref{fig:DisorderS}(b) with parameters consistent with the main text.
 The $f_{2}+f_3$ term involves (either virtual or resonant) pair excitation processes which produce a negative force that tends to reduce the gap.
The $f_1$ term is purely an intraband effect, where light excites thermally excited  quasiparticles to higher energies, producing a non-thermal distribution which gives a positive force that tends to enhance the gap, i.e., the physics behind the Eliashberg effect~\cite{PismaZhETF.11.186,klapwijk1977radiation,curtis2019cavity}. 
Relying on  thermally excited  quasiparticles, this contribution is exponentially suppressed at low temperatures but increases as $T\to T_{\text{c}}$.
In the dissipationless limit, meaing $T=0$, $\Omega < 2\Delta$ and $\eta=0$, one may verify analytically that \equa{EndMatterPforceResult} agrees with \equa{eqn:FP}, i.e.,
$
	F_{\text{P}}= \frac{i\omega}{c^2}  |A|^2 \partial_\phi \sigma
$.
It is evident from Fig.~\ref{fig:DisorderS}(b) that the dissipationless approximation (\equa{eqn:FP}) in the main text remains a good one as long as the  temperature is well below $T_{\text{c}}$.
As the temperature increases toward $T_{\text{c}}$,  $f_1$ gradually wins over $f_{2}+f_{3}$, resulting in a net positive force that enhances the gap, in agreement with the Eliashberg effect~\cite{PismaZhETF.11.186,klapwijk1977radiation,curtis2019cavity}.

Another finite temperature effect is the broadening of the cavity photon modes due to a nonzero $\text{Re}[\sigma]$ in the linear response. 
In a cavity pumped by external laser at the field strength $A_{\text{p}}$, the ponderomotive force acting on the gap is given by Eq.~(\ref{EndMatterPforceResult}) with $|A|^2$ replaced by $|\alpha'|^2 A^2_{\text{p}}$ and  $\alpha'(\Delta)$ from Eq.~(\ref{eq:alpha}). 
When $\text{Re}[\sigma]$ becomes nonzero, the superconductor introduces additional dissipation that broadens the resonance of $\alpha'$. 
This, in turn, broadens the ponderomotive potential step in \equa{eq:FpVp_step} and lowers the step-height $V_{\text{u}}$ similar to the 2DEG case, see SM Sec.~III~\cite{supp}.

\ifarXiv
    

\section*{Supplementary material}

sin $kh < 0$ in the phase transition region (see SI Sec. III). The critical field is thus larger than Eq. (10) by the inverse square root of this factor.

[63] M. Yarmohammadi, M. Bukov, and M. H. Kolodrubetz, Nonequilibrium phononic first-order phase transition in a driven fermion chain, Physical Review B **108**, L140305 (2023).
[64] M. Yarmohammadi, J. Sous, M. Bukov, and M. H. Kolodrubetz, Ultrafast dynamics of a fermion chain in a terahertz field-driven optical cavity, arXiv:2402.12591.
[65] Z. Sun, M. M. Fogler, D. N. Basov, and A. J. Millis, Collective modes and terahertz near-field response of superconductors, Phys. Rev. Res. **2**, 023413 (2020).
[66] D. C. Mattis and J. Bardeen, Theory of the Anomalous Skin Effect in Normal and Superconducting Metals, Phys. Rev. **111**, 412 (1958).
[67] P. W. Anderson, Theory of dirty superconductors, Journal of Physics and Chemistry of Solids **11**, 26 (1959).
[68] In the other limit of $\Delta_c \ll \Delta_0$, the criterion becomes $-V_0(\Delta_0) = V_u(E_c)$ and the critical field is $E_c \approx \Delta_0\sqrt{\nu k}$. However, note that if $\Delta_c$ is so small that $2\Delta_c < \omega$, the pump would cause pair breaking excitations and Eq. (5) no longer holds.
[69] M. Cardona, Electron Effective Masses of InAs and GaAs as a Function of Temperature and Doping, Phys. Rev. **121**, 752 (1961).
[70] M. Tinkham, Introduction to superconductivity (Courier Corporation, 2004).

# Supplemental Information for 'Universal phase transitions of matter induced in optically driven cavities'

**CONTENTS**

## I. ELECTROMAGNETIC MODELING AND OPTICAL COEFFICIENTS OF THE FABRY–PÉROT CAVITY

For a symmetric Fabry–Pérot cavity driven by incident pump light, the configurations of electric fields are shown schematically in Fig. S1(a). The EM boundary conditions are the top and bottom cavity mirrors are all encoded in their transmission ($T(\omega)$) and reflection ($R(\omega)$) coefficients, which satisfy $|T|^2 + |R|^2 = 1$ for dissipation-less mirrors.

We first derive the transmission coefficient $\alpha$ defined in Fig. S1(a). From the coefficients of the mirrors, one has

$$TE_\text{p} + RE_1 = E_2, \quad E_4 = E_2 e^{ik_z h}, \quad E_1 = RE_4 e^{ik_z h}. \tag{S1}$$

The ratio $\alpha(\omega)$ between electric field at the center of cavity and the incident pump field $E_p$ is thus found as:

$$E(z = h/2) = E_1 e^{-ik_z h/2} + E_2 e^{ik_z h/2} = \alpha E_\text{p}, \quad \alpha(\omega) = \frac{T(\omega) e^{i\frac{k_z h}{2}}}{1 - R(\omega) e^{ik_z h}}. \tag{S2}$$

Fig. S1(b) shows an oscillating in-plane dipole $\mathbf{P}_\perp e^{-i\omega t}$ placed on the center plane ($z = h/2$) of the cavity. The electric field is continuous at $z = h/2$ plane, so that $E_1 = E_2$. Ampère's circuital law applied to the $z = h/2$ plane leads to $-\frac{k_z}{\omega}(E_1 + E_2)(e^{-i\frac{k_z h}{2}} - R e^{i\frac{k_z h}{2}}) = \frac{4\pi}{c^2}\partial_t P$, which yields

$$E(z = h/2) = E_1 e^{-ik_z h/2} + RE_1 e^{ik_z h/2} = \beta \frac{\omega^2}{k_z c^2} P, \quad \beta(\omega) = 2i\pi \frac{1 + R(\omega) e^{ik_z h}}{1 - R(\omega) e^{ik_z h}}. \tag{S3}$$

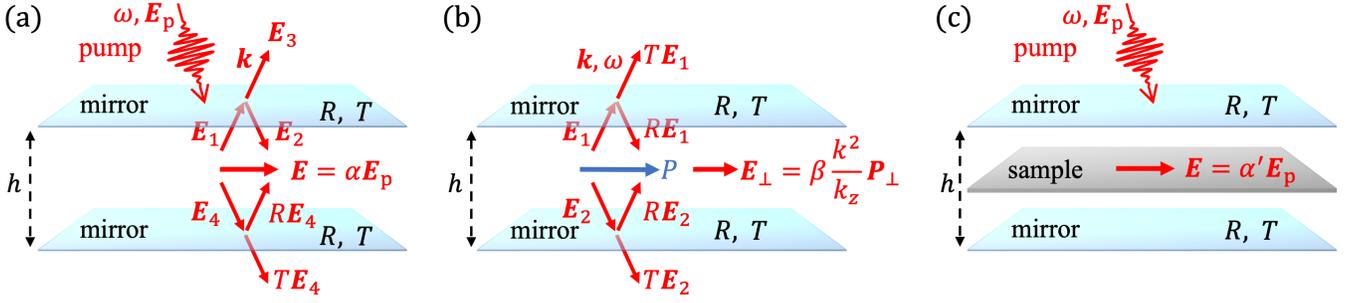

FIG. S1. (a) Illustration of the transmission coefficient $\alpha(\omega)$ of a bare Fabry–Pérot cavity with no sample inside it. It is defined as the ratio between the in-plane component of the electric field on the central plane and that of the incident light. (b) Illustration of the radiative coefficient $\beta(\omega)$ of the Fabry–Pérot cavity. The 2D electrical polarization $\mathbf{P}_\perp e^{i(\mathbf{q}_\perp \mathbf{r}_\perp - \omega t)}$ oscillating at frequency $\omega$ radiates light which, after repeated reflection by the mirrors, results in an electric field $\mathbf{E}$ on the plane of the polarization. The radiative coefficient $\beta(\omega)$ is defined as $E = \beta P_\perp k^2/k_z$ where $k_z$ is the out-of-plane momentum of the emitted light inside the cavity. The analysis is limited to TE mode here. In the limit of zero in-plane momentum $\mathbf{q}_\perp$, one has $E = \beta k P$. (c) Illustration of the transmission coefficient $\alpha'(\omega)$ of the Fabry–Pérot cavity with the sample inside it.

We are now ready to derive the transmission coefficient $\alpha'(\omega)$ of the Fabry–Pérot cavity with the sample inside it. In the configurations shown in Fig. S1(c), one has the relations $E(z = h/2) = \alpha E_\mathrm{p} + \beta \frac{\omega^2}{k_z c^2} P$ and $\partial_t P = \sigma E(z = h/2)$, which lead to

$$E(z = h/2) = \alpha' E_\mathrm{p}, \quad \alpha'(\omega) = \frac{\alpha}{1 - i\beta\sigma\omega/(k_z c^2)}\,. \tag{S4}$$

For normally incident light so that $k_z = \omega/c$, it reduces to Eq. 2 in the main text.

If the sample is placed not on the central plane in the cavity, the coefficients $\alpha(\omega)$, $\beta(\omega)$ will vary according to its location. A similar ponderomotive potential could be calculated using the position-dependent cavity coefficients. Generally, the sample can couple with both anti-node and node cavity photons with reasonable strength, so that a phase transition region exists near every cavity resonance modes at $kh = N\pi$. The critical pump field value will be larger since the field $A_s$ on the sample is smaller on resonance. Therefore, placing the sample on the central plane is the optimal choice, as this maximizes its coupling with certain cavity modes.

## II. DERIVATION OF THE PONDEROMOTIVE POTENTIAL

The ponderomotive force has been applied to a wide class of systems ranging from plasma [S42] to optical lattices [S43] and tweezers [S44], solid state materials [S42, S45–S53], and gravitational wave detection [S54, S55]. It was recently formalized for quantum many-body systems with the Keldysh path integral [S41]. There it is proved for two types of systems that the Taylor expansion coefficients of the ponderomotive potential $V_\mathrm{P} = \sum_n \chi^{(2n-1)}(\phi) E_\mathrm{p}^{2n}$ are set by the equilibrium response functions.

### A. Heuristic derivation of the step-like structure

To help the reader understand the step-like structure of the pondermotive potential, we start with a heuristic derivation using the model of a linearly driven harmonic oscillator [S41]:

$$L = \frac{1}{2}\left[-\dot{X}^2 + (\omega_0^2 + \phi)X^2\right] + E(t)X \tag{S5}$$

where $X$ is the displacement of the oscillator and $E(t) = E_0 e^{-i\omega t} + c.c.$ is the driving field. The slow degree of freedom $\phi$ couple to the oscillator by shifting its intrinsic frequency. Following the derivation of Eq. 11 in Ref. [S41], the total force on $\phi$ is simply the time-average of the force: $F = -\partial_\phi L = -\frac{1}{2}X^2$. The ponderomotive force is thus the driving field dependent part of its time average: $F_\mathrm{P} = -|\chi_R(\omega)E_0|^2$ where $\chi_R(\omega) = 1/(-\omega^2 - i\gamma\omega + \omega_0^2 + \phi)$ is the retarded response function of the oscillator. Therefore, one uncovers the Lorentzian structure of the ponderomotive force and the step-like structure of the ponderomotive potential:

$$F_\mathrm{P} = -E_0^2 \frac{1}{(\phi - \phi_c)^2 + \gamma_\phi^2}, \quad V_\mathrm{P} = \frac{E_0^2}{\gamma_\phi}\arctan\left[\frac{\phi - \phi_c}{\gamma_\phi}\right]\,. \tag{S6}$$

3where $\phi_c = \omega^2 - \omega_0^2$ and $\gamma_\phi = \gamma\omega$. Although derived from the classical equation of motion, this result is exact because of the exact quantum-classical correspondence of the retarded response of a harmonic oscillator, which does not have nonlinear responses. This force reflects a generic phenomenon of the ponderomotive force on a slow degree of freedom $\phi$ that affects a driven harmonic oscillator by shifting its eigen frequency: it points in the direction that red shifts the oscillator and results in a step-like potential.

### B. General expression of the ponderomotive force in an optically driven cavity

In the absence of dissipation from the sample, based on the conclusions in Ref. [S41], the Taylor expansion of the ponderomotive force in the light field on the plane of the sample is

$$F_{\rm P} = \sum_{n=1}^{\infty} \frac{-i}{\omega}|E(\omega)|^{2n} \partial_\phi \sigma^{(2n-1)} + \text{terms from } E(2\omega),\ E(3\omega), ...\ . \tag{S7}$$

Here $E(\omega) = i\omega A(\omega)/c$ is the electric field at frequency $\omega$ on the sample. Considering that there are in general optical nonlinearities from the sample and even the cavity itself, the electric field on the sample may also have frequency components at integer times the base frequency: $2\omega, 3\omega, ....$ Even the amplitude $E(\omega)$ at the fundamental frequency equal to the incident light may be modified by nonlinear optical effects. In some cases when there is spontaneous breaking of the discrete time translational symmetry, there may be other frequency components too, which we do not consider here.

### C. Universal step-like structure

In this section, we derive Eq. (5) in the main text which holds for a linear cavity, meaning neglecting the $n > 1$ terms in Eq. (S7) and neglecting the nonlinear optical corrections to $\alpha'$. The lowest order ponderomotive force in Eq. (S7) yields Eq. (4) which reads

$$F_{\rm P} = \frac{i}{c^2}\omega|\alpha'|^2 A_{\rm p}^2 \partial_\phi \sigma = \frac{i}{c^2}\omega A_{\rm p}^2 \left|\frac{\alpha}{1-i\beta\sigma/c}\right|^2 \partial_\phi \sigma = \frac{\frac{i}{c^2}\omega|\alpha|^2 A_{\rm p}^2 \partial_\phi \sigma}{(1-i\beta_1\sigma/c)^2 + (i\beta_2\sigma/c)^2} = \frac{\frac{i}{c^2}\omega|\alpha|^2 A_{\rm p}^2 \partial_\phi \sigma}{1 - 2i\beta_1\sigma/c + |\beta|^2(i\sigma)^2/c^2}. \tag{S8}$$

Here $\beta_1 = -\frac{4\pi R \sin kh}{1+R^2-2R\cos kh}$ and $\beta_2 = 2\pi \frac{1-R^2}{1+R^2-2R\cos kh}$ are the real and imaginary parts of $\beta$, and $|\beta|^2 = 4\pi^2 \frac{1+R^2+2R\cos kh}{1+R^2-2R\cos kh}$. We have ignored the phase of $R$, which can be absorbed in the $kh$ term. In the main text, $R$ is assumed to be close to $-1$. It is implicit that close to the pole of $F_{\rm P}$ as a function of $\phi$, it has the Lorentzian stucture of Eq. (S6).

Noting that $\alpha$ and $\beta$ are the bare cavity parameters that do not depend on $\phi$, the ponderomotive force can be formally integrated to yield the ponderomotive potential:

$$\begin{aligned} V_{\rm P} &= -\int d\phi F_{\rm P} = \frac{\omega}{c}|\alpha|^2 A_{\rm p}^2 \int \frac{d(-i\sigma/c)}{1+2\beta_1(-i\sigma/c)+|\beta|^2(i\sigma/c)^2} = \frac{V_{\rm u}}{\pi} \arctan\left(\frac{\beta_1 - i\sigma|\beta|^2/c}{\beta_2}\right), \\ V_{\rm u} &= \pi \frac{|\alpha|^2}{\beta_2}\frac{\omega}{c}A_{\rm p}^2 = \pi \frac{|T|^2}{2\pi(1-R^2)}\frac{\omega}{c}A_{\rm p}^2 = \frac{1}{2k}E_{\rm p}^2 = \frac{\lambda}{4\pi}E_{\rm p}^2. \end{aligned} \tag{S9}$$

Here we have used the relation $|\alpha|^2/\beta_2 = \frac{|T|^2}{2\pi(1-R^2)} = \frac{1}{2\pi}$ for dissipation-less mirrors. Note that $|\beta|^2/\beta_2 = 2\pi\frac{1+R^2+2R\cos kh}{1-R^2} = 2\pi\frac{1+R^2+2R\cos kh}{|T|^2}$. The potential drop $V_{\rm u}$ is independent on the transmission coefficient $T$ of the mirror because there is a cancellation from the line width $\gamma_{\rm r} = \omega|T|^2$ of the resonance.

## III. THE PONDEROMOTIVE FORCE FOR THE ELECTRON GAS

In this section, we derive the ponderomotive force for the 2D electron gas in the presence of a random disorder potential, resulting in a Drude optical conductivity of $\sigma(\omega) = i\frac{ne^2}{m(\omega+i\gamma)}$, where $\gamma$ is the scattering rate. The Lagrangian for the system is given by:

$$L = \int d^2 r\, \bar{\psi}\left[i\partial_t - \xi(\mathbf{p}+\mathbf{A}) - V_{\rm dis}\right]\psi + L_{\rm EM}[\mathbf{A}, \mathbf{A}_{\rm p}], \tag{S10}$$

4where $\langle V_{\text{dis}}(0) V_{\text{dis}}(r) \rangle = \delta(r) \frac{\gamma}{2\pi\nu}$ and $\nu$ is the density of states. Since we are primarily interested in the slow degrees of freedom of the system which is the the static electron density, we may integrate out the fast degrees of freedom, the high energy electron-hole excitations. To do this, one may introduce an energy cutoff $\Lambda$ for the electrons which is much smaller than the driving frequency $\omega_{\text{p}}$. If one integrates out the fast fermion fields $\psi_k$ with $\xi_k > \Lambda$, one obtains the Keldysh effective action [S56] for the slow fermion fields $\psi_s$ very close to the fermi surface:

$$S_{\text{eff}} = S_{\text{s}}[\psi_{\text{s}}, \mathbf{A}] + \int_{\omega > \Lambda} d\omega \mathbf{d}^2 r \frac{\sigma(\omega)}{i\omega} A^{\text{cl}}(\omega) A^{\text{q}}(-\omega) + S_{\text{EM}}[\mathbf{A}, \mathbf{A}_{\text{p}}]. \tag{S11}$$

Here the optical conductivity $\sigma(\omega)$ is contributed by the high energy electron-hole excitations with energies larger than $\Lambda$. However, since $\Lambda \ll \omega_{\text{p}}$, it is well approximated by the full optical conductivity: $\sigma(\omega) = i\frac{ne^2}{m(\omega + i\gamma)}$. The low energy electrons are now described by $S_{\text{s}}[\psi_{\text{s}}, \mathbf{A}]$, which are driven by the electrical field $\mathbf{A}(\omega_{\text{p}}) = \alpha'(\omega_{\text{p}}) \mathbf{A}_{\text{p}}(\omega_{\text{p}})$ with $\alpha'$ from Eq. (S4) and the optical conductivity being the Drude one. For a parabolic band $\xi(\mathbf{p}) = p^2/(2m) - \mu$, the diamagnetic term renders the ponderomotive force $F_{\text{P}} = -\frac{e^2}{2m} \langle A(t)^2 \rangle = -\frac{e^2}{m} |\alpha'|^2 A_{\text{p}}^2$.

The nonzero scattering rate leads to additional broadening of the cavity photon. Repeating the derivation in Eq. (S9), one finds that the step height of the ponderomotive potential is modified to:

$$V_{\text{u}} = \frac{\lambda}{4\pi} E_{\text{p}}^2 \times \frac{1 + (\frac{\gamma}{\omega})^2}{|1 + 2\frac{\gamma}{\omega} \frac{R}{|T|^2} \sin kh|}. \tag{S12}$$

If the phase of $R$ is taken into account and $kh$ is within the predicted phase transition range, one has $R \sin kh > 0$. Therefore, the potential drop $V_{\text{u}}$ is suppressed by impurity scattering by a factor of about $1/(1 + (2R \sin kh)\frac{\gamma}{\gamma_{\text{r}}})$.

In the clean limit of the 2D electron gas ($\sigma = i\frac{ne^2}{m\omega}$), the line-width of the cavity photon is determined by the radiative loss: $\gamma_{\text{r}} = \omega |T|^2$. From Eq. (S9), one has

$$\gamma_{\text{n}} = \frac{m\omega c}{2\pi e^2} \frac{|T|^2}{1 + R^2 + 2R \cos kh}, \quad n_{\text{c}} = \frac{m\omega c}{\pi e^2} \frac{R \sin kh}{1 + R^2 + 2R \cos kh} \tag{S13}$$

for Eq. 5 in the main text.

Lowest critical field— For pumping frequencies closely below the equilibrium cavity photon mode labeled by $\omega_0$ in Fig. 2(b) in the main text, the critical pump field $E_{\text{c}}$ could be very small. Unfortunately, due to the nonzero line-width $\gamma_{\text{n}}$ of the step potential, the critical pump field $E_{\text{c}}$ cannot be tuned to be exactly zero because the double-minimum structure in Fig. 1(d) of the main text will no longer exist if $n_0 - n_c \lesssim \gamma_{\text{n}}$, i.e., is the pumping frequency is too close to $\omega_0$. The lowest possible critical field can thus be estimated using Eq. 10 of the main text applied to the case $n_0 - n_c \sim \gamma_{\text{n}}$:

$$E_{\text{c}} \sim e\sqrt{\frac{2\pi}{C\lambda}} \gamma_{\text{n}} \sim e\sqrt{\frac{2\pi}{C\lambda}} \frac{m\omega c}{2\pi e^2} |T|^2 \sim \sqrt{\frac{d}{\lambda}} \frac{mc}{e} \gamma_{\text{r}}. \tag{S14}$$

For $m = 0.067 \, m_{\text{e}}$ corresponding to the electron mass in GaAs [S69], $d = \lambda$ and $\gamma_{\text{r}} = \omega_0 T^2 = 0.16 \, \text{meV}$, one has $E_{\text{c}} \sim 3 \times 10^5 \, \text{V/cm}$.

## IV. THE PONDEROMOTIVE FORCE FOR THE DISORDERED SUPERCONDUCTOR

At the linear response level, there is no dissipation for a sub-gap pump $\omega < 2\Delta$ in a superconductor at the zero temperature limit. The real part $\sigma_1(\omega)$ of the optical conductivity vanishes, while the imaginary part $\sigma_2(\omega)$ is given by [S70]:

$$\frac{\sigma_2(\omega, \Delta)}{\sigma_{\text{n}}} = \frac{1}{2}\left(1 + \frac{2\Delta}{\hbar\omega}\right) E(u') - \frac{1}{2}\left(1 - \frac{2\Delta}{\hbar\omega}\right) K(u'), \tag{S15}$$

where $\sigma_{\text{n}} = \frac{ne^2\tau}{m}$, $u' = \sqrt{1 - u^2}$, and $u = \frac{2\Delta - \hbar\omega}{2\Delta + \hbar\omega}$. Here, $K(u')$ and $E(u')$ are the complete elliptic integrals of the first and second kind, respectively. Using the results from Eq. (S9), the ponderomotive potential can be derived as

$$V_{\text{P}} = \frac{V_{\text{u}}}{\pi} \arctan\left(\frac{\beta_1 + \sigma_2(\omega, \Delta)|\beta|^2/c}{\beta_2}\right). \tag{S16}$$

The phase diagram in Fig. 3 of the main text is computed by minimizing Eq. 3 of the main text with the ponderomotive potential $V_{\text{P}}$ from Eq. (S16) and the optical conductivity $\sigma_2$ from Eq. (S15).

In the limit of $\omega \ll 2\Delta$, the imaginary part of optical conductivity can be approximated as a superfluid response $\sigma_2 \approx \frac{ne^2}{m}\frac{\pi\tau\Delta}{\hbar\omega}$. Accordingly, one has:

$$\gamma_\Delta = \frac{\hbar}{\pi\tau n}\frac{m\omega c}{2\pi e^2}\frac{|T|^2}{1+R^2+2R\cos kh}, \quad \Delta_c = \frac{\hbar}{\pi\tau n}\frac{m\omega c}{\pi e^2}\frac{R\sin kh}{1+R^2+2R\cos kh} \tag{S17}$$

for Eq. 5 in the main text. Compared to Eq. (S13) for the 2DEG, Eq. (S17) just has an extra prefactor.

<u>Lowest critical field</u>—Similar to the analysis for the 2D electron gas, the lowest possible critical field can be achieved by pumping at a frequency closely below the equilibrium cavity photon. The field could be estimated using the equation $E_c = (\Delta_0 - \Delta_c)\sqrt{2k\nu} = (\Delta_0 - \Delta_c)\sqrt{4\pi\nu/\lambda}$ in the main text applied to the case $\Delta_0 - \Delta_c \sim \gamma_\Delta$:

$$E_c \sim \gamma_\Delta \sqrt{4\pi\nu/\lambda} \sim \sqrt{4\pi\nu/\lambda}\frac{\hbar}{\pi\tau n}\frac{m\omega c}{2\pi e^2}|T|^2 \sim \frac{\sqrt{m/\lambda}}{\pi^2 ne\tau}\frac{mc}{e}\gamma_r, \tag{S18}$$

where we have used $\nu = \frac{m}{2\pi\hbar^2}$. Using the parameters for Fig. 3 in the main text, one has the shift of superconducting gap being $\gamma_\Delta \sim 0.2\,\text{K}$ and the critical field being $E_c \sim 0.5\,\text{V/cm}$.


\fi

\end{document}